\documentstyle[12pt]{article}
\begin{document}

\setlength{\textwidth}{15truecm}
\setlength{\textheight}{23cm}
\baselineskip=24pt


\bigskip

\centerline {\bf{VERTICAL DISTRIBUTION OF GALACTIC DISC STARS AND GAS}}

\centerline  {\bf {CONSTRAINED BY A MOLECULAR CLOUD COMPLEX }}

\medskip

\medskip

\medskip

\centerline {Chanda J. Jog and Chaitra A. Narayan}

\centerline  {Department of Physics, Indian Institute of Science}

\centerline {Bangalore 560012, India}

\centerline {email: cjjog@physics.iisc.ernet.in, chaitra@physics.iisc.ernet.in}

\newpage

\noindent{\bf ABSTRACT}

\noindent We investigate the dynamical effects of a molecular cloud  complex 
 with a mass $\sim 10^{7}M_{\odot}$ and a size $\sim$ a few  $100$ pc, on the
vertical distribution of stars and atomic hydrogen gas in a spiral galactic 
disc. Such massive complexes have now been observed in a number of spiral 
galaxies. The extended mass distribution in a complex, with an 
average mass density {\it six times higher} than the Oort limit, is shown to 
{\it dominate} the local gravitational field. This results in a significant 
re-distribution or clustering of the surrounding disc components towards the 
mid-plane, with a resulting decrease in their vertical scale-heights.

The modified, self-consistent stellar density distribution is
obtained by solving the combined Poisson equation and the force equation 
along the z-direction for an isothermal stellar disc on which the complex is 
imposed. The effect of the complex is strongest at its centre,
where the stellar mid-plane density increases
 by a factor of 2.6 and the vertical scale-height decreases  by a factor of 
 3.4 compared to the undisturbed stellar disc. A surprising
result is the large radial distance $\sim 500 $ pc from the
complex centre over which the complex influences the disc, this is due 
to the extended mass distribution in a complex. The complex has a comparable 
effect on the vertical distribution of the atomic hydrogen gas
in the galactic disc. This 
 `pinching' or constraining effect should 
be detectable in the nearby spiral galaxies, as for example has
been done for NGC 2403 (Sicking 1997).  Thus, the
gravitational field of a complex results in local
corrugations of the stellar and HI vertical scale-heights, and the
 galactic disc potential is highly non-uniform on scales of the 
inter-complex separation  of $\sim 1$ kpc.

Running Title: Vertical constraining by a cloud complex

\bigskip


\newpage

\noindent  {\bf 1. INTRODUCTION}

It is well known that the interstellar molecular hydrogen gas in the Milky Way
galaxy is contained in giant molecular clouds, with
 a typical radius of 20 pc and a mass of 
$\sim 5 \times 10^ 5 M_{\odot}$ each (Scoville \& Sanders 1987). It is
also established that the giant molecular clouds are further 
segregated in larger features which have been variously called as complexes 
or clusters or chains (Sanders et al. 1985, Rivolo, Solomon, \& Sanders 1986, 
Dame et al. 1987). These complexes are large, well-defined, discrete objects 
as seen by the closed contours of the CO distribution. Such a hierarchical 
distribution in complexes or  associations has also now been observed in
external galaxies such as M 51 (Lo et al. 1987, Rand \& Kulkarni 1990), 
M 83 (Lord \& Kenney 1991), and M 100 (Rand 1995), 
see Section 2 for details. The typical  molecular cloud complex 
 has a mass of  $\sim 10^7 M_{\odot}$ and a size of 
$\sim$ a few 100 pc, and a central total scale-height of about 120 pc.
Thus, the mass distribution in a complex is
elongated and it has an oblate spheroidal shape.

The average observed number density of hydrogen molecules in a
typical complex is about 16 $cm^{-3}$, which equals $ 1  M_{\odot} pc^{-3}$. 
We note that this is about {\it six times higher} 
than the dynamical or total mass density in the mid-plane of the
galactic disc in the solar neighbourhood as given by the Oort limit
of $0.15  M_{\odot} pc^{-3}$. (Oort 1960, Binney \& Tremaine
1987). This ratio may be even higher than six since the recent 
estimates based on the Hipparcos data (e.g.,
Binney 1997) give a smaller value of $0.076 M_{\odot} pc^{-3}$
for the Oort limit. Therefore, such a massive, extended mass 
distribution in the form of a molecular cloud complex in a
galactic disc would {\it dominate} the local gravitational field. 
Thus the stars and the atomic hydrogen gas in the surrounding
galactic disc would respond to the additional and dominant gravitational 
field due to the complex, and tend to get clustered towards 
the mid-plane, with a resulting decrease in 
their vertical scale-heights. We investigate this 
effect quantitatively, and show that the above intuitive picture is confirmed 
by our calculations.  

A similar clustering effect will also be seen around an individual 
giant molecular cloud, though with a much smaller magnitude and over a smaller 
spatial range than for the case of a complex.

We show that the gravitational field of a complex results in a local 
variation of the stellar and HI vertical scale-heights, and the net disc 
potential is locally non-uniform. This would be of interest in the general
theoretical studies of galactic disc dynamics since the usual assumption of 
uniform and smooth mass distribution or potential is shown to no longer hold 
good.

A similar problem involving the effect of dark halo on the vertical 
distribution of the embedded disc matter was studied by Bahcall (1984). 
However, in that case the halo acts as a perturbation in the plane and hence a 
perturbation analysis was used. In our work, 
on the other hand, the complex dominates the gravitational field in the
disc region around it.

 In another analogous problem, a planar disc response to the
perturbation due to an imposed massive cloud taken to be a point mass was 
treated by Julian \& Toomre (1966), whereas we have treated the effect  
 on the vertical disc distribution due to a spatially extended mass 
 distribution in a complex which dominates the local gravitaional field. 
 Further, we argue 
that the complex has a stronger effect on the vertical than the planar disc 
distribution (Section~5). 

In Section~2, the observed physical parameters for the typical
giant molecular cloud complex, and also the
choice of the galactic disc potential used are described. The
formulation of the equations for the self-consistent response of
a disc to the imposed complex are given in Section~3. The results on the 
vertical constraining effects of a complex on stars and HI gas  and
a comparison with observational data are given in Section~4 .  A discussion 
of a few general points are given in Section~5, and Section~6 contains a 
brief summary of our conclusions.

\noindent {\bf {2. PARAMETERS FOR COMPLEX AND GALACTIC DISC}}

\noindent {\bf {2.1 Parameters for a cloud complex}}

Observations of a number of spiral galaxies have now clearly
established the existence of giant molecular cloud complexes or
associations in our Galaxy (see \S 1), and also in external galaxies: for 
example as in M51 (Lo et al. 1987, Rand \& Kulkarni 1990), M83 
(Lord \& Kenney 1991), M 31 (Allen \& Lequex 1993), SMC (Rubio, 
Lequex, \& Boulanger 1993), M100 (Rand 1995), NGC 4414 (Sakamoto 1996), 
NGC 5055 (Thornley \& Mundy 1997), M81 (Brouillet et al. 1998), and for 
the central regions of 20 nearby spiral galaxies (Sakamoto  et al. 1999) -
some of which overlap with the galaxies listed above.   
Thus, such large gas features appear to be common in spiral galaxies.
		
The molecular cloud complexes in these galaxies are observed to cover a 
wide range of masses from $10^6$ to $10^8 M_{\odot}$,
with a {\it typical mass} of $10^7  M_{\odot} $ and a {\it
typical size} of a few hundred pc.  These complexes are shown to be 
spatially coherent features (Rivolo et al. 1986), and are statistically 
significant and are not mere superpositions or artifacts of the low spatial  
resolution of the observations (Rivolo et al. 1986, Rand \& Kulkarni 1990, 
Sakamoto 1996).

Gravitationally bound complexes with the above typical values are shown to 
occur naturally via induced gas instabilities in a
gravitationally coupled star-gas two-fluid galactic disc (Jog \&
Solomon 1984). From the observed star complexes, Efremov (1995)
has argued that gas `superclouds' of $\sim 10^7 M_{\odot} $ occur in all 
spiral galaxies. Rand \& Kulkarni (1990) show for M 51 that features above 
a few times $10^7 M_{\odot}$ may not be gravitationally bound. Hence for 
our work, we use a typical complex of mass equal to $10^7 M_{\odot}$ and a 
radius of 200 pc. We assume a maximum height at the centre of the complex to 
be equal to 120 pc which is the FWHM as observed for the molecular gas in the 
galactic disc (Sanders, Solomon, \& Scoville 1984, Clemens et al. 1986). 
Further, the mass distribution inside a complex is assumed to be uniform for 
simplicity.
This gives a mean molecular hydrogen gas density, $n$, in a typical complex to 
be equal to 15.8 $H_2 cm^{-3}$. The mass density in a complex, 
$\rho_{complex}$, assuming a 10 \% gas number fraction in helium, is equal 
to $2.8 m_H n$ where $m_H$ is the atomic mass of hydrogen.
For the typical complex, $\rho_{complex}$ is equal to 1
$M_{\odot} pc^{-3}$.

This is a conservative choice because a complex more massive
than this will have a stronger effect on the stars but since it
is not bound the effect will be short-lived, hence we do not consider 
complexes more massive than $10^7 M_{\odot}$. Also for these more massive 
and extended complexes, the 
effect of the differential rotation in the disc cannot be neglected.	

It has been claimed by some authors that the molecular gas distribution is 
hierarchical and may be denoted as a fractal, however this covers largest 
entities of sizes $\sim 100$ pc which are self-gravitating (Falgarone, 
Phillips, \& Walker 1991). The fractal description therefore does not seem 
to cover the larger size bound complexes that are observed in a number of 
galaxies.

The complex is taken to be co-planar with the galactic disc, and
$z$ is along normal to the galactic plane, with $z=0$ denoting
the galactic mid-plane. The radial distance $r$ is measured from
the complex centre,  taken to be at $r = 0$. Here $r$
and $z$ are the cylindrical polar co-ordinates.
The mass distribution in a cloud complex is  that of an oblate
spheroid, and the force per unit mass due to it along $z$ is taken from 
Schmidt (1956) who had obtained it in the context of
galaxy mass models:

$$ (K_z)_{complex} \: = \: 4 \pi G \: {\rho}_{complex} \: {\epsilon}^{-3} \: 
    ( 1 - {\epsilon}^2 )^{1/2} \: z \: (tan \: \beta \: - \: \beta) 
         \eqno (1) $$
    
\noindent where $\beta$ is given within and outside the
ellipsoid respectively by:

$$ sin \beta \: = \: \epsilon  \eqno (2) $$

$$ r^2 \: sin^2 \: \beta \: + \: z^2 \: tan^2 \: \beta \: = \:
a^2 \: {\epsilon}^2 .  \eqno (3) $$

\noindent Here $\epsilon$ is the eccentricity of the oblate
spheroid and is equal to $\epsilon =  ( 1 - c^2 / a^2 )^{1/2}$
where $c$ and $a$ are the semi-minor and semi-major axes along the $z$ and 
the radial directions respectively. For a typical complex, $c = 60 $
pc and $ a = 200 $ pc as discussed above, so that $\epsilon$ = 0.95.

\noindent {\bf {2.2 Parameters for the galactic disc}}

We have assumed the galactic disc to consist of a single isothermal stellar 
component with the density distribution along $z$ satisfying the $sech^2$ law 
(Spitzer 1942), rather than assuming a more realistic, multi-component 
galactic mass model say by Carlberg \& Innanen (1981). This was done for 
simplicity since it captures the main $z$ distribution of the galactic disc, 
and because it allows us to clearly illustrate the physical effect of the 
imposed complex on the stellar disc. 

Further we assume for simplicity that the stellar distribution remains
isothermal even in presence of a complex, with the z-velocity
dispersion, $<{(v_z)_s}^2>^{1/2}$, the same as observed in the undisturbed 
disc, namely 15 km s$^{-1}$ (Binney  \& Merrifield 1998). For an isothermal, 
one-component, self-gravitating
stellar disc, the density distribution, $ {\rho}_s$, and the
force per unit mass, $(K_z)_s$, are obtained by solving
the force equation along $z$ along with the Poisson equation (Spitzer 1942,
Rohlfs 1977) and are given respectively by:

$$ \rho_s \: = \: \frac {<{(v_z)_s}^2>}{2 \pi G {z_0}^2} \: \:
      sech^2 \: (\frac {z}{z_0}) \eqno (4) $$

$$  (K_z)_s \: = \: \frac {- 2 \: <{(v_z)_s}^2>}{z_0} \: tanh \:
  \frac  {z}{z_0}          \eqno (5) $$

\noindent where

$$ z_0 \: = \: \left [ \frac {<{(v_z)_s}^2>}{2 \pi G \rho_0 }
    \right ] ^{1/2}        \eqno (6) $$

\noindent here $z_0$ is taken to be a measure of the thickness of the
undisturbed disc but is not identically equal to the vertical
scale-height $h_{1/2}$ (or HWHM) which we will use in Section 4
and later sections. For small $z/z_0$, the density distribution is that of 
a gaussian.

To get a quantitative feel for the effect of the complex, in  Fig. 1 we plot 
the magnitude of the z-force per unit mass due to a complex (eq.[1]) and that 
due to the undisturbed stellar disc without the complex (eq.[5]) versus $z$ at
 the complex centre ($r = 0$).
It is interesting that upto $z \leq$ 200  pc, the force due to
the complex
dominates over that due to the disc. This ratio a is maximum, equal
to 9.5, at the outer edge of the complex,  where the
force due to the complex is a maximum. Because of the large value of
this ratio, we are justified in assuming that the complex itself is not 
affected by the stellar disc.		

Fig. 2 shows the ratio of the force per unit mass along the
$z$ direction due to the complex to that due to the undisturbed stellar
disc versus $z$ at different radial distances from the centre of the complex. 
The ratio is large $\sim 9.5$  at the mid-plane over $r=0$ to $r=200$
pc and decreases gradually with the radial and vertical
distance, and indicates the large spatial range over which the complex affects
the disc. This large range is confirmed for the self-consistent
disc response in Section~4.

\noindent {\bf {3. FORMULATION OF EQUATIONS}	}

We first formulate the equations for the self-consistent response of
the stellar disc to the imposed complex potential and obtain the
modified stellar density distribution. 

\noindent {\bf 3.1 {Effect on Stars}}
	
We obtain the density distribution along the $z$ direction at a
given radial distance, $r$, from the centre of the complex, so
that we look at the distribution along a normal cut across a complex. 
The equations to be solved to study the modified stellar mass
density distribution, $\rho_s$, in the presence of an imposed
cloud complex are the force equation along the $z$ direction and
the joint Poisson equation for the stars and the complex :

$$ \frac {<{(v_z)_s} ^2>}{\rho_s} \: \frac {d \rho_s}{dz} \: = \:
      (K_z)_s \: + \: (K_z)_{complex}  \eqno (7) $$
      
\noindent and,

$$ \frac {d (K_z)_s}{d z} \: + \: \frac {d (K_z)_{complex}}{d z}
  \: + \: \frac {d (K_r)_{complex}}{d r} \: + \: \frac {(K_r)_{complex}}
    {r}    \: = \: - 4 \pi G (\rho_s \: + \: \rho_{complex} )
                \eqno (8)  $$

\noindent where $(K_z)_s$ and $ (K_z)_{complex}$ are the
z-components of the force per unit mass due to the modified
stellar distribution and the complex respectively, and $(K_r)_{complex}$
is the $r$ component of the force per unit mass due to the complex. 
Since the stellar disc is thin compared to its radial extent, only the 
$z$ component of the Laplacian 
for the stars needs to be retained on the l.h.s. of the Poisson
equation, as usual (see e.g. Binney \& Tremaine 1987, pg. 199).
For the complex, on the other hand, both $z$ and $r$ terms are required
since the $z$ and $r$ dimensions are comparable. 
Because of the azimuthal symmetry w.r.t. the complex centre, the $\phi$ term
for both stars and the complex is zero.

The complex is assumed to be unaffected by the disc (Section~2.2), hence the 
joint Poisson equation (eq.[8]) effectively reduces to that for the 
re-distributed 
stellar case alone. Thus, on combining the above two coupled first-order 
linear differential equations in $\rho_s$ and $(K_z)_s$ (eq.[7]-[8]), we get 
the following linear, second-order differential equation in $\rho_s$ :

$$ \frac {d}{d z} \: \left ( \frac {<{(v_z)_s}^2>}{\rho_s} \: 
    \frac {d \rho_s}{dz} \right ) \: = \: - 4 \pi G \rho_s \:
    + \:  \frac {d (K_z)_{complex}}{d z}  \eqno (9) $$

\noindent where the term due to the complex on the
r.h.s. of equation (9) is obtained starting from equation (1).

The above formulation effectively assumes the complex to be stationary
in the disc and long-lived. This is justified since the response time of a
star given by $2 \pi / \nu = 6 \times 10^7 $ yr, where $\nu$ is
the vertical frequency (Binney \& Tremaine 1987), is comparable to the
crossing time of the complex in the plane, and it is smaller
than the life-time of the complex which is gravitationally bound.

\noindent {\bf 3.2 {Effect on Atomic Hydrogen Gas}}

We next consider the effect of a molecular cloud complex and the
modified stellar distribution on
the atomic hydrogen gas in a galactic disc,  also taking account of the 
self-gravity of the atomic gas. The atomic hydrogen gas is assumed to be 
isothermal as is known from gas dynamics, with the z-velocity
dispersion, $<{(v_z)_g}^2>^{1/2}$ = 8 km s$^{-1}$  (Spitzer 1978).
The modified distribution of the gas is given by:

$$ \frac {d}{d z} \: \left ( \frac {<{(v_z)_g}^2>}{\rho_g} \: 
    \frac {d \rho_g}{dz} \right ) \: = \: - 4 \pi G \rho_s \: 
    - 4 \pi G \rho_g \:
    + \:  \frac {d (K_z)_{complex}}{d z}  \eqno (10) $$

\noindent where $\rho_s$ denotes the re-distributed stellar density
obtained from equation (9). The gas has a negligible effect on
stars in the presence of a complex, hence we do not include it
in the study of stellar distribution in Section 3.1.
		
\noindent {4. RESULTS}

\noindent {\bf 4.1 {Effect on Stars}}

First, we consider the effect of a molecular cloud complex on stars,
and obtain the modified stellar density distribution.
The problem studied in the present paper is analogous to that of the
re-distribution of charges in the presence of an external
imposed electric field, that is commonly studied in the electromagnetic 
theory (e.g., Jackson 1976). This analogy was first saught to help solve
equation (9) by assuming that the stellar distribution
reverts back to the undisturbed value at large $z$.
However, we find that even if we start at very large z values of $\sim 2 $ 
kpc, the disc still retains the memory of the initial solution, and hence one
cannot exploit this analogy to solve this problem. This is
because while 2 kpc is a large distance physically for the disc
being studied, mathematically it is still not large enough for
this approach to work.

Instead, the second-order, linear differential equation in
$\rho_s$ (eq. [9]) is solved as an initial value problem from inside-out 
numerically  using the fourth-order Runge-Kutta method (Press
et al. 1986). The initial condition of $d \rho_{s} / d z = 0$ at $z=0$ is used
since that would be true for any realistic mass distribution in a disc. 
The value of the re-distributed density $\rho_{s}$ at the mid-plane ($z = 0$) 
is not known a priori. We constrain this by
assuming that the total column density of the 
undisturbed stellar disc, $\Sigma_s$, remains constant even in the presence
of a complex. Thus, the complex is assumed to re-destribute the
stars only along $z$ at a given radius, and thus we have
solved this 1-D problem. The radial force due to
a complex can be neglected as discussed in Section 5. 

The complex centre is taken to be located at the
galactocentric radius of 8.5 kpc, in the solar neighbourhood. Here, the 
unperturbed central, mid-plane stellar density is known from observations 
to be  $0.07 M_{\odot}$ pc$^{-3}$ (Binney \& Tremaine 1987). 
The equation (9) with zero contribution from the complex is 
 solved and this gives the variation with $z$ in the
self-gravitating, undisturbed stellar mass density (Fig. 3, dashed line).
This shows a $sech^2 z$ distribution as expected for an
undisturbed disc (eq.[4]). The net column or surface density for
stars, $\Sigma_s$, at a given radius is 
obtained by summing up the area under the curve and doubling it and this is 
found to be 47.6  $M_{\odot}$ pc$^{-2}$.
Assuming the stellar surface density at a given radius to remain constant 
when the complex is imposed,
equation (9) is solved to obtain the modified or re-distributed stellar
density distribution, which is also shown in Fig. 3 (solid
line). {\it The complex strongly alters the vertical distribution of the 
stellar disc.}   First, 
the central stellar density at $z=0$ increases by a factor of 2.6 to be
equal to $0.19 M_{\odot}$ pc$^{-3}$. Second, the vertical distribution still
satisfies a $sech^2$ law but now the scale-height $h_{1/2}$ (HWHM) of the
distribution has decreased by a factor of 3.4 to be equal to 88 pc. This 
`pinching' or constraining effect confirms the intuitive picture given in 
Section~1.

This procedure is repeated at different radii from the complex
centre to get the net 2-D re-distribution of the stars,
where the constant net stellar column density of 47.6 $M_{\odot}$ pc$^{-2}$
is used as explained above.  In Fig. 4,  the resulting modified vertical
scale-height, $h_{1/2}$, is plotted as a function of
radial distance from the complex centre (solid line). Also shown
is the constant undisturbed scale-height (dashed line). The most 
striking result is the large radial distance from the complex centre
over which the complex influences the stellar distribution. This interesting 
and somewhat unexpected result is due to the extended mass distribution in a 
complex. The trough representing the scale-height is  broad so that even at a 
radial distance of 500 pc from the complex centre, the  
vertical scale-height decrease is still  $8 \% $. The scale-height 
asymptotically reverts back to the undisturbed value of 299 pc only beyond 
$r = $ 1 kpc. The broad trough makes it more likely that this
effect would be detected in observations (see Section 4.4). Thus, the 
resulting stellar scale-height distribution is locally 
corrugated, see Section~4.4 for a comparison with observations.

In Fig. 5, we show a contour diagram  for the iso-density
contours of the modified stellar distribution to show another aspect of the 
constraining effect due to the complex. In the presence of a complex, a 
higher range of density values is covered over a smaller
z-range than for the undisturbed disc. As in the variation of scale-height 
(Fig. 4), the density also asymptotically reverts back to the 
undisturbed value of $0.035 M_{\odot}$ pc$^{-3}$ at the
scale-height = 299 pc, only
beyond $r =$ 1 kpc, as expected since the scale-height and
density variations are inter-related.  An interesting point is
that the density contours are clearly curved
around the complex, thus highlighting the important effect of
the complex in re-distributing the stellar matter around it.

\noindent {\bf 4.2 {Effect on stars: Parameter Study}}

We study the effect of a complex over the
parameter space covering a wide range of realistic values for
the complexes and the stellar disc density. The results from
this study are given in Table 1. Columns 2-4 give the input
parameters for the complex, namely the semi-major axis $a$,
the semi-minor axis $c$, and the molecular hydrogen number
density $n $. Column 5 gives the total stellar disc
surface density $\Sigma_s$. Columns 6 and 7 give the resulting mid-plane
 density and the vertical scale-height $h_{1/2}$ for the modified stellar
 distribution.

Case 1 is the typical case studied in Section 4.1 (Fig. 4), which
is used as a benchmark with which we compare the results from the other cases.
Case 2 with double the total disc surface density of case 1 represents
the peak of the molecular ring region at 6 kpc. This follows
from the radial dependence of surface density for an
exponential disc ($\propto exp [-r / r_{exp}]$) with a disc
scale-length, $r_{exp}$, of 
3.5 kpc as in the Galaxy (Binney \& Tremaine 1987).
 Here the unperturbed central density is 
0.28 $M_{\odot}$ pc$^{-3}$   and the scale-height is 150 pc as obtained
by solving equation (9) with zero contribution from the complex. 
The effect of a complex is smaller in this case, with the mid-plane
density increasing and the scale-height decreasing only by a
factor of 2 - see the last
two columns, case 2 in Table 1.
The trough showing the effect of a complex is shallower and
less broad in this case because a higher density
disc is less disturbed by the same complex.

The converse result would be seen for a lower stellar disc surface
density where the effect of a complex would be stronger. This
could be applicable for the few, less-massive cloud
complexes observed in the outer Galaxy (Digel, Thaddeus, \& Bally 1990).

A smaller value of the complex number density of 10 $H_2$ $cm^{-3}$
results in a smaller increase in the mid-plane density to  
0.14 $M_{\odot} pc^{-3}$
at $r=0$, and a larger central scale-height of 119 pc, as expected because 
the trough in  the scale-height is shallower (case 3). A complex with a
smaller  radius $a$ (case 4), or with a smaller
 height $c$ (case 5),  with 
the same density as in case 1, also results in an effect of
smaller magnitude as expected.

\noindent {\bf 4.3 {Effect on Atomic Hydrogen Gas}}

The modified HI distribution is obtained by solving 
equation (10) where $\rho_s$ is obtained as a solution to
equation (9) (see Section 3.2).  The atomic hydrogen gas 
surface density is taken to be constant at 5 $M_{\odot}$
pc$^{-2}$ as observed in the
inner disc (Scoville \& Sanders 1987). This, plus the 
assumption of the gradient of the gas density to be zero at $z$ = 0, as done 
before for the stellar case (Section 4.1),  
allows us to solve equation (10).
The re-distributed gas density at the mid-plane at the central radius of
the complex is higher by a factor of 3.2 and the vertical 
gas scale-height is lower by a factor of 3.3 compared to the case
without the complex. Fig. 6 shows the vertical scale-height for
HI as a function of radius. The  vertical 
constraining of HI gas as a function of input parameters would show a trend 
similar to that for stars as studied in Section 4.2.

\noindent {\bf 4.4 {Comparison with Observations}}

\noindent 1. {\it The `pinching' or the constraining effect:}  The `pinching' 
or the constraining effect of a complex on the z-distribution
   of stars and gas (Sections 4.1 - 4.3) is strong in magnitude and it 
affects stars and gas upto a large radial distance of $\sim 500 $ pc from the 
complex centre (Fig. 4). Nevertheless it could be missed 
unless it is looked for carefully. This is because in an edge-on
galaxy like say NGC 891, the
cumulative effect over many such complexes along a line-of-sight
would average out and hence cannot be seen easily, and in a face-on
galaxy the spatial z-distribution cannot be studied directly. Further, for 
the external
galaxies, the uncertainty in the determination of the inclination angle and 
the ubiquitous presence of warps (Burton 1992) and also the flaring of gas can
affect the determination of scale-height.

   However, if one were to carefully analyze the information for the 
column density $N (l,b,v)$ where $l,b,v$ are the galactic
longitude, latitude and the line-of-sight velocity respectively, and model 
the data cube say as one does to study the spiral structure parameters, 
then in principle, the z scale-height variation predicted in our paper 
can be verified from observations for a nearby, inclined
galaxy. This result for variation is in fact confirmed 
from the  high-resolution HI 
mapping  of NGC 2403 by Sicking (1997) who
found that the vertical scale-height of HI shows variations on a scale 
of about 100 arc sec, which corresponds to about 1.3 kpc for the distance 
of 3.5 Mpc for this galaxy. This agrees with a total of $\sim$ 
 1 kpc radial range affected by a complex as shown in our paper.
Note that this is a late-type Sc galaxy with  a large amount of molecular
gas detected in it (Young et al. 1995).

For a more detailed spatial comparison,  we need a resolution of about 
2 arc sec corresponding to 100 pc at say a distance of 10 Mpc.
The current HI surveys of galaxies are typically at 20 arc sec 
(e.g. Binney \& Merrifield 1998, pg. 488). Thus only in our 
Galaxy the HI data have sufficient resolution to allow a more 
detailed study of the variation in scale-height due to a complex that 
we predict. We will attempt this in a future paper.

   A recent, interesting paper has shown how the
 HI scale-height for face-on galaxies such as
LMC may be determined using the technique of spectral correlation 
function (Padoan et al. 2001). Here the scale-height
can be determined upto spatial scales of the resolution for their
data, namely, 20 pc. If this technique could be applied to HI data 
from other galaxies in which molecular cloud complexes have been 
observed (e.g., M 51, M 81), then it would give the variation in 
vertical scale-height around a complex. Note that since the effect 
of a complex is felt over 500 pc, and the above galaxies are a few 
Mpc away, the resolution required to study this variation is about 
20 arc second which is exactly within the resolution of large-scale 
HI surveys of these galaxies.

   Our paper also predicts the constraining effect for the old
stars in the disc, which can be checked in the near-IR data on 
galaxies since that traces old stars. Unfortunately, in the
literature on external galaxies, the data is averaged over radius  to give only
a typical vertical scale-length, as e.g. in the large sample of 486 
galaxies studied by Ma et al. (1997). For our 
Galaxy also, the near-IR study by Freudenreich (1998)
gives an average vertical scale-height, whereas the best fit to the
near-IR data by  Kent, Dame, \& Fazio (1991) requires 
a constant scale-height upto a radius of 5 kpc and radially increasing 
beyond that. This agrees with the flaring behaviour expected if we take 
account of the average effect of the complexes in the inner Galaxy
(see Section 4.4, point 2).

Thus, the observational verification of our results on the
vertical constraining due to a complex is within
the present observational capabilities, especially for a nearby,
slightly tilted galaxy.

\noindent 2. {\it Flaring of HI :}  
 The molecular gas and hence the cloud complexes are
located  within the solar circle (Scoville \& Sanders 1987).
Thus the vertical constraining of the atomic hydrogen gas due to
the molecular cloud complexes would abruptly stop at this radius and
this could naturally explain the long-standing puzzle of why the atomic 
hydrogen gas shows flaring exactly outside of the solar position
(e.g., Burton \& te Lintel Hekkert 1986, Wouterloot et al. 1990). The details
of this will be checked in a future paper that involves taking account of the
effect of many such complexes in the disc (also see Sections
4.5, and 5).

\noindent 3. {\it Corrugation of the vertical scale-heights:}  We have shown 
that a complex would result in the local corrugation of the vertical
scale-heights of stars and gas.
The previous studies of corrugation in the galactic 
disc typically refer to the variation in the mid-plane ($<z>$)
with radius of HI (e.g. Lockman 1977, Spicker \& Feitzinger 1986),
or the old stars (Florido et al. 1991), the physical origin of which 
is not yet understood, and we do not study it here.
The stellar corrugation is color-dependent which Florido et al. (1991)
attribute to star formation. We note that the local corrugation of the 
vertical scale-heights 
obtained in the present paper is an additional complexity, seen
on top of the corrugation of the mid-plane, and has to be reckoned with 
while studying the $z$ distribution in a spiral galactic disc.
If the underlying molecular gas mid-plane is corrugated as it is in
our Galaxy (Sanders et al. 1984), then the corrugation of scale-heights
predicted here is a local phenomenon seen around a complex. It 
is probably easiest to detect this effect around a complex in
nearby galaxies such as M 51 or M 81.

\noindent {\bf 4.5 {Implications for Galactic Dynamics}}

The massive, extended cloud complex and the resulting clustering of stellar 
disc mass 
around it means that the mass distribution and hence the potential of a 
galactic disc in a spiral galaxy is not uniform. The galactic disc potential 
is  $\it non-uniform$ on the scales of a typical observed separation 
between complexes $\sim 1$ kpc (Section 5). This could be important in the 
general theoretical 
studies of galactic disc dynamics in spiral galaxies. For example, this may
contribute to the observed increase with age (Wielen 1977) of the stellar 
random velocity dispersion.  The increase in stellar dispersion
is a classic problem, and in the past it has been 
attributed as arising due to scattering of stars off molecular clouds
(Spitzer \& Schawrzschild 1951, Lacey 1984), or due to local spiral features
(Barbanis \& Woltjer 1967), or as a combination of both these processes
(Jenkins  1992). The cloud complexes present a 
new source of non-uniform potential on a spatial  scale that lies
between these two cases studied earlier. In a future paper, we will study the
heating caused by the cloud complexes, and thus check
the importance of the integrated dynamical effect of cloud complexes in a
galactic disc.

The potential distribution resulting due to several such complexes 
would have the appearance of the "muffin-tin"
potential that is used to study solids in condensed matter
physics (e.g., Ashcroft \& Mermin 1976).
It would be very useful if future observations give the mass spectrum
and the distances between such complexes in a spiral galaxy. This will decide 
if the influence of the various `potential wells' located at the
complex centres are distinct or they merge together (also see Section
5). This will be studied in the context of flaring in a future
paper (see Section 4.4).

\noindent {\bf { 5. DISCUSSION}}

 (1).  The total molecular gas mass in the disc of our Galaxy is
equal to $2 \times 10^9 M_{\odot}$, most of which is observed to be in the 
inner disc (Scoville \& Sanders 1987).
 Assuming that all the molecular gas mass in the disc is in cloud complexes, 
there are about $\sim 200$ such complexes in the inner Galaxy and the
typical separation between two neighbouring complexes can be
calculated to be $\sim 1 $ kpc. Since each complex can affect the
distribution of stars and gas upto  500 pc and beyond from its centre
on either side (see Figs. 4 and 6), hence the dynamical effect of the 
complexes taken together can effectively `cover' or affect most 
of the disc. To address this question in detail, it would be necessary to have
observational determination of the mass spectrum of the
complexes, and the fraction of molecular hydrogen gas mass that exists in
complexes in a galactic disc.

 (2).  We have assumed the net stellar disc column or surface density, 
 $\Sigma_s$,  at a
given radius $r$ to remain constant in the presence of a complex and have used
 this constraint to
obtain the re-distributed stellar disc density (Section 4.1).
The complex will also exert a radial force on the
surrounding disc components. However, the ratio of the radial to the
$z$ component of the force due to a complex is $< $ 1 (Schmidt 1956), except
near the mid-plane at the outer boundary of a complex.
Further, from the radial force equation (e.g., Binney \& Tremaine 1987), it 
can be shown that the ratio of the change in density along $r$ to that along 
$z$ will be proportional to the ratio of vertical to the radial mean square of
the random velocity, which is observed to be 1/4 (see Binney \&
Merrifield 1998, Chap. 10). Therefore, we 
can ignore the transfer or re-distribution of matter along the radial 
direction as compared to that along $z$, hence the assumption of constant 
surface density is valid.

  (3).   We have assumed for simplicity that the stellar disc surface
density is constant across the complex. In a realistic galactic disc 
the mid-plane density increases at lower galactic radii. Hence the net effect 
of complex will be lower at the inner boundary of the complex than at 
the highest radius (see Section 4.2). For example, an exponential disc
mass distribution with a scale-length of $3.5$ kpc as observed for the Galaxy
(Section 4.2), results in a $ 8 \% $ variation in the disc
surface density between the inner and the outer edges of a
complex. This results in a $ 7 \% $ difference in the scale-heights
at the inner and outer boundaries of the complex. Thus the pinching effect 
would show a radial asymmetry, and 
 the net scale-height distribution would be horn-shaped with the broad-end of 
 the horn at a smaller radius.

\noindent {\bf { 6. CONCLUSIONS}}

We have studied the dynamical effects of a massive, extended giant molecular 
cloud complex, with a mass $\sim 10^7 M_{\odot}$ and a size $\sim$ a few
hundred pc, on the z-distribution of stars and atomic 
hydrogen gas in a spiral galactic disc. The main results obtained are:

(1).  Physically, a superposition of a co-planar
potential due to the complex on the stellar disc re-distributes the stars 
so that the stellar density at the mid-plane is higher, and
correspondingly, the scale-height of the stellar distribution is
lower. The constraining effect of the complex is highest at its central 
radius, where the mid-plane stellar density is shown to increase 
by a factor of 2.6, and the vertical scale-height is shown to decrease by a 
factor of 3.4. The complex significantly affects the disc upto a 
large radial distance of 500 pc from its centre, and the 
undisturbed distribution is reached slowly beyond $\sim $ 1 kpc. This is
a direct consequence of the extended mass distribution in a complex.
The complex is shown to have a comparable effect on the
vertical distribution of the atomic hydrogen gas in the galactic disc.

(2). The verification of the `pinching' or constraining effect predicted 
around a cloud
complex is within the present observational capabilities, especially for a 
nearby, slightly tilted galaxy. In fact, this result has been
confirmed from the observations of HI in NGC 2403 (Sicking 1997). Thus, the 
gravitational field of a complex results in local corrugations of the 
vertical scale-heights of the stellar and HI  distributions in a galactic disc.

(3). The massive complex, and the resulting significant clustering of
stellar mass around it, means that the  
gravitational potential in a typical spiral galactic disc is distinctly 
non-uniform on scales of the inter-complex seperation
of $\sim 1$ kpc. This would be of 
interest in the general theoretical studies of galactic disc dynamics since 
the usual assumption of a smooth and uniform disc potential is
shown to be not valid.   

(4).  The observed onset of flaring of HI gas 
outside of the solar circle is probably due to the observed drop
in the molecular gas content beyond this radius, where the vertical 
constraining due to the cloud complexes would not be operative.
This will be studied in detail in a future paper.

(5). This paper has highlighted the importance of the
dynamical effects of cloud complexes on a galactic disc.
We stress that the
complexes should be included as a major structural component in
the study of a spiral galaxy. 
Future observations that give the mass spectrum and distances between
such complexes in a spiral galaxy are needed, these will help in deciding
how far the effect of the nearby complexes overlaps.

\medskip

\noindent {\bf {ACKNOWLEDGEMENTS}}   
   
We would like to thank D. Chakraborty for useful discussions
during the early stages of this work, and  R. Nityananda,
H. Bhatt, and T. Prabhu for useful comments. We would like to
thank the referee, P. Grosbol, for detailed and constructive
comments which clarified several points in the 
paper, and improved the discussion on the comparison of our results 
with observations.

\newpage

\noindent{\bf REFERENCES}

\bigskip

\noindent Allen, R.J., \& Lequex, J. 1993, ApJ, 410, L15

\noindent Ashcroft, N.W.,  \& Mermin, N.D. , 1976, Solid State Physics
     (New York: Holt, Rinehart, \& Winston)  

\noindent Bahcall, J. 1984, ApJ, 276, 156

\noindent Barbanis, B., \& Woltjer, L. 1967, ApJ, 150, 461

\noindent Binney, J. 1997, in Highlights of Astronomy, ed. J. Andersen, 
     11A, 412.

\noindent Binney, J., \& Merrifield, M. 1998, Galactic Astronomy,
           (Princeton : Princeton University Press)
     
\noindent Binney, J.,  \& Tremaine, S. 1987, Galactic Dynamics,
           (Princeton : Princeton University Press)

\noindent Brouillet, N., Kaufman, M., Combes, F., Baudry, A., \&
              Bash, F.  1998, A \& A, 333, 92

\noindent Burton, W.B. 1992, in The Galactic Interstellar
  Medium, Saas-Fee Advanced course 21, Eds., D. Pfenniger, \&
  P. Bartholdi (Berlin: Springer- Verlag)

\noindent Burton, W.B., \& te Lintel Hekkert, P. 1986, A \& A Suppl., 
    65, 427
    
\noindent Carlberg, R.G., \& Innanen, K.A. 1987, AJ, 94, 666

\noindent Clemens, D.P., Sanders, D.B., Scoville, N.Z., \&
 Solomon, P.M. 1986, ApJS, 60, 297

\noindent Dame, T.M.  et al. 1987, ApJ, 322, 706

\noindent Digel, S., Thaddeus, P., \& Bally, J.  1990, ApJ, 357, L29

\noindent Efremov, Y.N., 1995, AJ, 110, 2757

\noindent Falgarone, E., Phillips, T.G., \& Walker, C.K. 1991,
 ApJ, 378, 186

\noindent Florido, E., Battaner, E., Sanchez-Saavedra, M.L.,   Prieto, M., 
 \& Mediavilla, E.   1991, MNRAS, 251, 193

\noindent Freudenreich, H.T. 1998, ApJ, 492, 495

\noindent Jackson, J.D.  1976, Classical Electrodynamics (New York: 
     John Wiley)

\noindent Jenkins, A. 1992, MNRAS, 257, 620

\noindent Jog, C.J., \& Solomon, P.M. 1984, ApJ, 276, 127

\noindent Julian, W.H., \& Toomre, A. 1966, ApJ, 146, 810

\noindent Kent, S.M., Dame, T.M., \& Fazio, G. 1991, ApJ, 378, 131 

\noindent Lacey, C.G. 1984, MNRAS, 208, 687

\noindent Lo, K.Y., Ball, R., Masson, C.R., Phillips, T.G.,
Scott, S., \& Woody, D.P. 1987, ApJ, 317, L63

\noindent Lockman, F.J. 1977, AJ, 82, 408

\noindent Lord, S.D., \& Kenney, J.D.P. 1991, ApJ, 381, 130

\noindent Ma, J., Peng, Q.-H., Chen, R., \& Ji, Z.-H.  1997,
  A \& AS, 126, 503  

\noindent Oort, J.H. 1960, Bull.Astron.Inst. Netherlands, 15, 45

\noindent Padoan, P., Kim, S., Goodman, A., \& Staveley-Smith, L.
    2001, submitted to ApJL, (astro-ph/0103251).

\noindent Press, W.H., Flannery, B.P., Teukolsky, S.A., \&
  Vetterling, W.T. 1986, Numerical Recipes (Cambridge: Cambridge
  Univ. Press), chap. 6.

\noindent Rand, R.J.  1995, AJ, 109, 2444

\noindent Rand, R. J., \& Kulkarni, S.R. 1990, ApJ, 349, L43

\noindent Rivolo, A.R., Solomon, P.M., \& Sanders, D.B. 1986, ApJ, 301, L 19

\noindent Rohlfs, K. 1977, Lectures on Density Wave Theory,
  (Berlin: Springer- Verlag)

\noindent Rubio, M., Lequex, J., \& Boulanger, F. 1993, A \& A,
   271, 9

\noindent Sakamoto, K. 1996, ApJ, 471, 173

\noindent Sakamoto, K., Okumura., S.K., Ishizuki, S., \&
 Scoville, N.Z. 1999, ApJS, 124, 403 

\noindent  Sanders, D.B., Clemens, D.P., Scoville, N.Z., \&
 Solomon, P.M. 1985, in The Milky Way Galaxy, IAU Symposium 106, eds. H. van
 Woerden et al. (Dordrecht: Reidel), 329

\noindent Sanders, D.B., Solomon, P.M., \& Scoville, N.Z. 1984,
  ApJ, 276, 182

\noindent Schmidt, M. 1956, BAN, 13, 15

\noindent Scoville, N.Z., \& Sanders, D.B. 1987, in Interstellar Processes, 
  eds. D.J. Hollenbach \& H.A. Thronson (Dordrecht: Reidel), 21

\noindent Sicking, F.J. 1997, Ph.D. thesis, University of Groningen.

\noindent Spicker, J., \& Feitzinger, J.V. 1986, A \& A, 163, 43

\noindent Spitzer, L. 1942, ApJ, 95, 329

\noindent Spitzer, L. 1978, Physical Processes in the
   Interstellar Medium (New York: John Wiley)

\noindent Spitzer, L., \& Schwarzschild, M. 1951, ApJ, 114, 385

\noindent Thornley, M.D., \& Mundy, L.G. 1997, ApJ, 490, 682

\noindent Wielen, R. 1977, A \& A, 60, 263

\noindent Wouterloot, J.G.A., Brand, J., Burton, W.B., Kwee,
 K.K. 1990, A \& A, 230, 21

\noindent Young, J.S. 1995, ApJS, 98, 219

\newpage

\centerline{\bf FIGURE LEGENDS}

\noindent {\bf FIG. 1.} $\:\: - \:\:$  A plot of
$\vert (K_z)_{complex} \vert $, the magnitude of the force per unit mass along 
$z$ due to a complex 
and that due to the undisturbed stellar disc versus $z$, the
distance from the mid-plane, at the 
complex centre. Upto $z \leq$ 200 pc, the force due to the complex
dominates over that due to the disc. The ratio is a maximum, equal
to 9.5, at the outer edge of the complex (z = 60 pc).

\noindent {\bf FIG. 2.} $\:\: - \:\:$  A plot of 
$(K_z)_{complex}$/$(K_z)_{s}$, the ratio of the force due to the complex to 
that due to the undisturbed stellar disc, versus $z$ at different 
radial distances from the complex centre, in steps of 50 pc. The curves for 
$r=0, 200, $ and $250$ pc only are labeled for clarity.
The ratio is large $\sim 9.5$  at the mid-plane over $r=0$ to $r=200$
pc and decreases gradually with the radial and vertical distance.

\noindent {\bf FIG. 3.} $\:\: - \:\:$ A plot of the modified stellar 
disc density in the presence of a cloud complex
versus $z$, the distance from the mid-plane (solid line) at the
centre of the complex, and also the
self-gravitating, undisturbed stellar disc density without the complex (dashed
line). The modified mid-plane density is 2.6 times higher and the vertical 
scale-height is smaller by a factor of 3.4 than the undisturbed case, showing 
the strong `pinching' or constraining effect due to the complex.

\noindent {\bf FIG. 4.} $\:\: - \:\:$ A plot of the vertical 
scale-height $h_{1/2}$ (HWHM) for the modified stellar
distribution versus $r$, the radial distance from the complex centre 
(solid line), and the constant scale-height for the undisturbed
stellar disc without the complex (dashed line).
The vertical constraining by the complex is significant
over a large radial distance $\sim 500$ pc from its centre,  this is due 
to the extended mass distribution in a complex.

\noindent {\bf FIG. 5.} $\:\: - \:\:$ A contour diagram for the constant
density contours of the modified stellar distribution in the $r,
z$ plane in units of $M_{\odot} pc^{-3}$.  A higher range of stellar density 
values (upto 0.18 $M_{\odot} pc^{-3}$) is covered over a  smaller z-range than
 for the undisturbed disc, this is due to the constraining 
 effect of the complex. The contours are clearly curved
around the complex, thus highlighting the important effect of
the complex in re-distributing the stellar matter around it.

\noindent {\bf FIG. 6.} $\:\: - \:\:$ A plot of the vertical scale-height 
 for the modified HI gas distribution versus $r$, the radial
distance from the complex centre (solid line), and the constant 
gas scale-height for an undisturbed disc without the complex  (dashed line). 
The constraining 
effect of the complex on the gas is similar to that for the stellar 
case (see Fig. 4).

\pagebreak
\eject
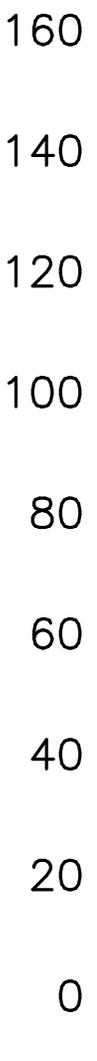
\end{document}